\documentclass[12pt,onecolumn,showpacs,amsmath,amssymb]{revtex4}

\begin{document}

\title{Nonlinear dynamics of the interface of dielectric liquids
in a strong electric field: Reduced equations of motion}

\author{Nikolay M. Zubarev}
\email{nick@ami.uran.ru}

\affiliation{Institute of Electrophysics, Ural Branch, Russian
Academy of Sciences,\\ 106 Amundsen Street, 620016 Ekaterinburg, Russia}

\begin{abstract}

The evolution of the interface between two ideal dielectric liquids in a 
strong vertical electric field is studied. It is found that a particular
flow regime, for which the velocity potential and the electric field
potential are linearly dependent functions, is possible if the ratio
of the permittivities of liquids is inversely proportional to the ratio of
their densities. The corresponding reduced equations for interface motion
are derived. In the limit of small density ratio, these equations coincide
with the well-known equations describing the Laplacian growth.  

\end{abstract}


\maketitle

It is well known that the flat interface of two dielectric
liquids is unstable in a sufficiently strong vertical electric field. The
dispersion relation for the surface waves has the following form
\cite{mel,mel1}:
$$
\omega^2=\frac{\rho_1-\rho_2}{\rho_1+\rho_2}\,gk
-\frac{E_1E_2(\varepsilon_1-\varepsilon_2)^2}{4\pi(\rho_1+\rho_2)
(\varepsilon_1+\varepsilon_2)}\,k^2
+\frac{\alpha}{\rho_1+\rho_2}\,k^3,
$$
where $k$ is the wave number, $\omega$ is the frequency, $g$ is the
acceleration of gravity, $\alpha$ is the surface tension coefficient,
$\rho_1$ and $\rho_2$ are the mass densities of lower and of upper
liquids ($\rho_1>\rho_2$), $\varepsilon_1$ and $\varepsilon_2$ are the
dielectric constants of fluids. The external electric field strengths
under and above the interface, $E_1$ and $E_2$, are related by the
expression
\begin{equation}
\varepsilon_1 E_1=\varepsilon_2 E_2.
\label{0}
\end{equation}
It is seen from the dispersion relation that, if the electric field is
sufficiently strong, 
$$
E_1E_2\gg
\frac{\varepsilon_1+\varepsilon_2}{(\varepsilon_1-\varepsilon_2)^2}
\sqrt{g\alpha(\rho_1\!-\!\rho_2)},
$$
the second term in right-hand side of the dispersion relation dominates
for the waves with wave numbers in the range 
$$
\frac{g(\varepsilon_1+\varepsilon_2)(\rho_1-\rho_2)}
{E_1E_2(\varepsilon_1-\varepsilon_2)^2}\ll
k\ll
\frac{E_1E_2(\varepsilon_1-\varepsilon_2)^2}
{\alpha(\varepsilon_1+\varepsilon_2)}.
$$
Then $\omega^2\propto k^2$ and, hence, we can separate the dispersion
relation into two branches 
\begin{equation}
\omega^{(\pm)}=\pm ick, \qquad
c^2=\frac{E_1E_2(\varepsilon_1-\varepsilon_2)^2}{4\pi(\rho_1+\rho_2)
(\varepsilon_1+\varepsilon_2)}.
\label{1}
\end{equation}
For one branch, small periodic perturbations of the surface increase
exponentially with the characteristic times $(ck)^{-1}$, while, for the
other branch, these perturbations attenuate. In such a situation, 
we can restrict our consideration to the increasing branch
$\omega^{(+)}=+ick$, that essentially simplifies the problem of
describing the evolution of the interface at the linear
stage of the development of instability. The buildup of perturbations of
the surface inevitably transforms the system to a state in which its
evolution is determined by nonlinear processes. Then, in the general case,
splitting into the branches becomes impossible.

In this paper we will show that, for the particular case 
$\varepsilon_1\rho_1=\varepsilon_2\rho_2$, we can extract the separate
branches from the equations of motion. This makes it possible to reduce by
half the number of equations required for describing the evolution of the
boundary. The reduced equations coincide with the well-known equations
describing the Laplacian growth in the limit of small ratio of liquid
densities. An important point is that the Laplacian growth equations 
not only define a subclass of particular solutions of the problem, but
they also describe the asymptotic behavior of the system.

It should be noted that the behavior of the interface of two fluids in
normal electric or magnetic field (these problems are similar from the
mathematical point of view) is usually investigated in the
quasi-monochromatic approximation (see \cite{moh,sin,cal,elf} and the 
references therein). This approach allows one to obtain immediately an
equation for the complex amplitude of surface waves. However, the
applicability of such an equation is limited by the condition of the
smallness of the slopes of the surface. The development of instability
can violate this condition. In the strong-field limit, the approach
developed in the present work provides a way of studying the interface
behavior at essentially nonlinear stages of instability development.

Consider the evolution of the interface of two ideal liquids of
infinite depth in an external vertical electric field. In the unperturbed
state, the boundary of the liquid is a flat horizontal surface.
Let the $z$ axis of the Cartesian coordinate system is normal to
the unperturbed interface. The function $\eta(x,y,t)$ specifies the shape
of the deformed boundary, i.e., the liquids occupy the regions
$z<\eta(x,y,t)$ and $z>\eta(x,y,t)$, respectively. It is convenient for
the subsequent analysis to choose an origin of coordinates so that the
level of liquids is determined by the expression $z=-vt$. In other words,
the origin moves with respect to the interface at a certain constant
velocity $v$. 

Let us assume that the motion of both liquids is potential.
The velocity potentials for incompressible liquids $\Phi_1$ and $\Phi_2$
satisfy the Laplace equations, 
\begin{equation}
\nabla^2\Phi_1=0, \qquad \nabla^2\Phi_2=0,
\label{2}
\end{equation}
with the following conditions at the boundary and at infinity:
\begin{gather}
\rho_1\left[\frac{\partial\Phi_1}{\partial t}+
\frac{(\nabla\Phi_1)^2}{2}\right]
-\rho_2\left[\frac{\partial\Phi_2}{\partial t}+
\frac{(\nabla\Phi_2)^2}{2}\right]
=\frac{\varepsilon_1-\varepsilon_2}{8\pi}\,
(\nabla\varphi_1\cdot\nabla\varphi_2), \quad z=\eta(x,y,t),
\label{3}
\\
\frac{\partial\Phi_1}{\partial n}=\frac{\partial\Phi_2}{\partial n},
\qquad z=\eta(x,y,t),
\label{4}
\\
\Phi_1\to -vz, \qquad z\to-\infty,
\label{5}
\\
\Phi_2\to -vz, \qquad z\to+\infty,
\label{6}
\end{gather}
where $\varphi_1$ and $\varphi_2$ are the electric-field potentials in and
above the liquid, and $\partial/\partial n$ denotes the derivative along
the normal to the interface. The expression on the right-hand side of the 
dynamic boundary condition (nonstationary Bernoulli equation) is
responsible for the electrostatic pressure at the interface between two
ideal dielectric liquids in the absence of free electric charges
\cite{lanlif}. The evolution of the interface is determined by the
kinematic relation, 
\begin{equation}
\frac{\partial\eta}{\partial t}=\frac{\partial\Phi_1}{\partial z}-
(\nabla_{\!\!\bot}\eta\cdot\nabla_{\!\!\bot}\Phi_1), \qquad z=\eta(x,y,t).
\label{7}
\end{equation}

The electric potentials $\varphi_1$ and $\varphi_2$ satisfy the Laplace
equations,
\begin{equation}
\nabla^2\varphi_1=0, \qquad \nabla^2\varphi_2=0.
\label{8}
\end{equation}
Since the electric field potential and normal component of the
displacement vector have to be continuous at the interface, we should add
the following conditions at the boundary:
\begin{gather}
\varphi_1=\varphi_2, \qquad z=\eta(x,y,t),
\label{9}
\\
\varepsilon_1\frac{\partial\varphi_1}{\partial n}=
\varepsilon_2\frac{\partial\varphi_2}{\partial n}, \qquad z=\eta(x,y,t).
\label{10}
\end{gather}
The system of equations is closed by the condition of the
electric field uniformity at an infinite distance from the surface:
\begin{gather}
\varphi_1\to -E_1z, \qquad z\to-\infty,
\label{11}
\\
\varphi_2\to -E_2z, \qquad z\to+\infty.
\label{12}
\end{gather}

Let us show that a flow regime, wherein the harmonic potentials of
velocity and of electric field are linearly dependent functions, is
possible for certain relations between the problem parameters. Suppose 
that 
\begin{equation}
\varphi_1=a\,\Phi_1(4\pi\rho_1/\varepsilon_1)^{1/2}, \quad
\varphi_2=b\,\Phi_2(4\pi\rho_2/\varepsilon_2)^{1/2},
\label{13}
\end{equation}
where $a$ and $b$ are unknown constants. It is necessary to verify that
the initial equations of motion (\ref{2})--(\ref{12}) are compatible with
these relations. Substituting them into (\ref{3}) and
(\ref{9})--(\ref{12}), we obtain
\begin{gather}
\rho_1\!\left[\frac{\partial\Phi_1}{\partial t}+
\frac{(\nabla\Phi_1)^2}{2}\right]\!
-\!\rho_2\!\left[\frac{\partial\Phi_2}{\partial t}+
\frac{(\nabla\Phi_2)^2}{2}\right]\!
=\!ab(\varepsilon_1\!-\!\varepsilon_2)
\frac{\sqrt{\rho_1\rho_2}}{\sqrt{\varepsilon_1\varepsilon_2}}\,
\frac{(\nabla\Phi_1\!\cdot\!\nabla\Phi_2)}{2},
\quad z=\eta(x,y,t),
\label{14}
\\
\Phi_1 a\,(4\pi\rho_1/\varepsilon_1)^{1/2}=
\Phi_2 b\,(4\pi\rho_2/\varepsilon_2)^{1/2}, \qquad z=\eta(x,y,t),
\label{15}
\\
\frac{\partial\Phi_1}{\partial n}\,a\,(\rho_1\varepsilon_1)^{1/2}=
\frac{\partial\Phi_2}{\partial n}\,b\,(\rho_2\varepsilon_2)^{1/2},
\qquad z=\eta(x,y,t),
\label{16}
\\
\Phi_1\to -z\,E_1a^{-1}(4\pi\rho_1/\varepsilon_1)^{-1/2},
\qquad z\to-\infty,
\label{17}
\\
\Phi_2\to -z\,E_2b^{-1}(4\pi\rho_2/\varepsilon_2)^{-1/2},
\qquad z\to+\infty.
\label{18}
\end{gather}
For the system of equations (\ref{2}), (\ref{4})--(\ref{7}) and
(\ref{14})--(\ref{18}) to be compatible (it is overdetermined in the
general case), the conditions (\ref{4})--(\ref{6}) must coincide with the
conditions (\ref{16})--(\ref{18}), and the condition (\ref{7}) must
coincide with the condition (\ref{14}). 

It is apparent that the conditions (\ref{4}) and (\ref{16}) coincide if  
\begin{equation}
a(\rho_1\varepsilon_1)^{1/2}=b(\rho_2\varepsilon_2)^{1/2}.
\label{19}
\end{equation}
In view of Eqs.~(\ref{0}) and (\ref{19}), the conditions at infinity
(\ref{5}), (\ref{6}) and (\ref{17}), (\ref{18}) are consistent if the
auxiliary parameter $v$ takes the following value: 
$$
v=a^{-1}v_0, \qquad v_0=E_1(4\pi\rho_1/\varepsilon_1)^{-1/2}>0.
$$
 
Finally, we consider the condition under which the dynamic (\ref{14}) and
kinematic (\ref{7}) relations coincide. Let us write Eq.~(\ref{7}) in the
form which does not contain function $\eta$ explicitly. With the help of
the formula (\ref{19}), the boundary condition (\ref{15}) can be rewritten
as follows: 
$$
{\varepsilon_1}^{-1}\Phi_1={\varepsilon_2}^{-1}\Phi_2,
\qquad z=\eta(x,y,t).
$$
Differentiating this expression with respect to time or spatial variables,
we arrive at
\begin{gather*}
\frac{\partial\eta}{\partial t}\cdot
\left[\varepsilon\,\frac{\partial\Phi_1}{\partial z}-
\frac{\partial\Phi_2}{\partial z}\right]_{z=\eta}\!\!
=-\left[\varepsilon\,\frac{\partial\Phi_1}{\partial t}-
\frac{\partial\Phi_2}{\partial t}\right]_{z=\eta}\!,
\\
\nabla_{\!\!\bot}\eta\cdot\left[
\varepsilon\,\frac{\partial\Phi_1}{\partial z}-
\frac{\partial\Phi_2}{\partial z}\right]_{z=\eta}\!\!
=-\left[\varepsilon\,\nabla_{\!\!\bot}\Phi_1-
\nabla_{\!\!\bot}\Phi_2\right]_{z=\eta},
\end{gather*}
where $\varepsilon=\varepsilon_2/\varepsilon_1$ is the ratio of the 
permittivities. These relations allow us to eliminate $\eta$ from 
Eq.~(\ref{7}). We obtain from the kinematic boundary condition:
\begin{equation}
\varepsilon\,\frac{\partial\Phi_1}{\partial t}-
\frac{\partial\Phi_2}{\partial t}=-
\varepsilon\,(\nabla\Phi_1)^2+
(\nabla\Phi_2\cdot\nabla\Phi_1),
\qquad z=\eta(x,y,t)
\label{21}
\end{equation}
Decomposing the velocities of fluids into the normal
($\partial\Phi/\partial n$) and tangential ($\partial\Phi/\partial\tau$)
components in Eqs.~(\ref{14}) and (\ref{21}), and taking into account
Eqs.~(\ref{4}), (\ref{15}) and (\ref{19}), we get
\begin{gather*}
\frac{\partial\Phi_1}{\partial t}-
\rho\,\frac{\partial\Phi_2}{\partial t}+
\frac{(1\!-\!\rho\!-\!a^2\varepsilon^{-1}\!+\!a^2)}{2}
\left[\frac{\partial\Phi_1}{\partial n}\right]^2\!\!+
\frac{(1\!-\!\rho\varepsilon^2\!-\!a^2\!+\!a^2\varepsilon)}{2}
\left[\frac{\partial\Phi_1}{\partial\tau}\right]^2\!\!=0,
\quad z=\eta(x,y,t)
\\
\frac{\partial\Phi_1}{\partial t}-
\varepsilon^{-1}\frac{\partial\Phi_2}{\partial t}+(1-\varepsilon^{-1})
\left[\frac{\partial\Phi_1}{\partial n}\right]^2\!\!=0,
\qquad z=\eta(x,y,t),
\end{gather*}
where $\rho=\rho_2/\rho_1$. Clearly, these expressions can coincide only
if the following conditions hold: 
\begin{gather*}
\rho=\varepsilon^{-1}, \qquad
1-\rho\varepsilon^2-a^2+a^2\varepsilon=0, 
\\
1-\rho-a^2\varepsilon^{-1}+a^2=2-2\varepsilon^{-1}.
\end{gather*}
From this it is inferred that the equations are compatible
provided that $a^2=1$ and also
\begin{equation}
\varepsilon_1\rho_1=\varepsilon_2\rho_2.
\label{22}
\end{equation}
The equation for the parameter $a$ has two roots, $a^{(\pm)}=\pm 1$,
corresponding to different branches of solutions.

Thus, we have proved that the functional relation (\ref{13}) can be
compatible with the equations of motion if the condition (\ref{22})
is valid. The corresponding flow regime is described by the following
equations: 
\begin{gather}
\nabla^2\Phi_1=0, \qquad \nabla^2\Phi_2=0,
\label{23}
\\
\frac{\partial\eta}{\partial t}=\frac{\partial\Phi_1}{\partial n}\,
\sqrt{1+(\nabla_{\!\!\bot}\eta)^2},
\qquad z=\eta(x,y,t),
\label{24}
\\
\frac{\partial\Phi_1}{\partial n}=\frac{\partial\Phi_2}{\partial n},
\qquad z=\eta(x,y,t),
\label{25}
\\
\rho_1\Phi_1=\rho_2\Phi_2, \qquad z=\eta(x,y,t),
\label{26}
\\
\Phi_1\to -a^{(\pm)}v_0z, \qquad z\to-\infty,
\label{27}
\\
\Phi_2\to -a^{(\pm)}v_0z, \qquad z\to+\infty.
\label{28}
\end{gather}
The reduction of the initial equations (\ref{2})--(\ref{12}) to
Eqs.~(\ref{23})--(\ref{28}) significantly simplifies the analysis of the 
interface motion. As will be discussed below, in the formal limit
$\rho_2/\rho_1\to 0$, these equations describe the so-called 
Laplacian growth.

Let us find the dispersion relation for Eqs.~(\ref{23})--(\ref{28}).
We will seek a solution in the form 
\begin{gather*}
\Phi_1=c_1e^{i(kx-\omega t)}e^{+k(z+a^{(\pm)}v_0t)}-a^{(\pm)}v_0z-{v_0}^2t,
\\
\Phi_2=c_2e^{i(kx-\omega t)}e^{-k(z+a^{(\pm)}v_0t)}-a^{(\pm)}v_0z-{v_0}^2t,
\\
\eta=c_3e^{i(kx-\omega t)}-a^{(\pm)}v_0t,
\end{gather*}
where $c_1$, $c_2$, and $c_1$ are small constants. These expressions
correspond to a small-amplitude sinusoidal deformation of the initially
plane liquid-liquid interface. After simple transformations, we obtain the
following relation between the frequency $\omega$ and the wave number $k$:
$$
\omega=ia^{(\pm)}\,\frac{\rho_1-\rho_2}{\rho_1+\rho_2}\,v_0k=
a^{(\pm)}\frac{i\,\sqrt{\varepsilon_1}\,(\rho_1-\rho_2)}
{\sqrt{4\pi\rho_1}\,(\rho_1+\rho_2)}\,E_1k.
$$
It can be seen that, for the branch $a=a^{(+)}=+1$, initial perturbation
will increase and, for $a=a^{(-)}=-1$, it will attenuate. It should be
noted that, with regard to Eqs.~(\ref{0}) and (\ref{22}), this expression
coincides with the expression (\ref{1}) specifying different branches 
of the dispersion relation for the unreduced equations of motion. 

Thus, if the condition (\ref{22}) is satisfied, the separation of two
branches corresponding to solutions increasing and decreasing with time
is possible not only in the linearized equations, but also in the initial
nonlinear equations (\ref{2})--(\ref{12}).

The question arises as to whether the flow regime under consideration
is stable. In other words, whether or not Eqs.~(\ref{23})--(\ref{28}) 
describe the large-time asymptotic behavior of the system. 
Stability of the increasing branch is evident at linear stages of the
interface evolution, when the linearized equations of motion can be split
into two independent systems. At the nonlinear stages, the equations do
not split completely, and the stability problem becomes nontrivial.

It turns out that the stability can be proved in the limiting case
$\rho_1\gg\rho_2$ and $\varepsilon_1\ll \varepsilon_2$ (the condition
(\ref{22}) can be violated). Then the evolution of the interface will be
governed by the influence of the lower liquid. The equations determining
the interface motion become: 
\begin{gather*}
\nabla^2\Phi_1=0, \qquad \nabla^2\varphi_1=0,
\\
\frac{\partial\Phi_1}{\partial t}+
\frac{(\nabla\Phi_1)^2}{2}+
\frac{\varepsilon_1(\nabla\varphi_1)^2}{8\pi\rho_1}=0,
\qquad z=\eta(x,y,t),
\\
\frac{\partial\varphi_1}{\partial t}+(\nabla\Phi_1\cdot\nabla\varphi_1)=0,
\qquad z=\eta(x,y,t),
\\
\varphi_1=0, \qquad z=\eta(x,y,t),
\\
\Phi_1\to -vz, \qquad \qquad z\to-\infty,
\\
\varphi_1\to -E_1z, \qquad z\to-\infty,
\end{gather*}
where the kinematic boundary condition is given in the implicit form.
If we introduce a pair of auxiliary potentials, 
$$
\Psi^{(\pm)}=
2^{-1}\Phi_1\pm(16\pi\rho_1/\varepsilon_1)^{-1/2}\varphi_1,
$$
these equations can be rewritten in the following symmetric form (compare
with Refs.~\cite{zub,zub1}):
\begin{gather}
\nabla^2\Psi^{(\pm)}=0,
\label{29}
\\
\frac{\partial\Psi^{(\pm)}}{\partial t}+
\left(\nabla\Psi^{(\pm)}\right)^2=0,
\qquad z=\eta(x,y,t),
\label{30}
\\
\Psi^{(+)}=\Psi^{(-)}, \qquad z=\eta(x,y,t),
\label{31}
\\
\Psi^{(+)}\to -v_0z, \qquad z\to-\infty,
\label{32}
\\
\Psi^{(-)}\to 0 \qquad z\to-\infty.
\label{33}
\end{gather}
Here, we set $a=+1$ and, as a consequence, $v=v_0$.

One can readily see that these equations are compatible with the condition
$\Psi^{(-)}=0$, which corresponds to the situation of interest,
where the velocity potential and the electric field potential are
functionally related. For $\Psi^{(-)}=0$, the set of equations
(\ref{29})--(\ref{33}) reduces to 
\begin{gather}
\nabla^2\Psi^{(+)}=0,
\label{34}
\\
\frac{\partial\eta}{\partial t}=\frac{\partial\Psi^{(+)}}{\partial n}\,
\sqrt{1+(\nabla_{\!\!\bot}\eta)^2},
\qquad z=\eta(x,y,t),
\label{35}
\\
\Psi^{(+)}=0, \qquad z=\eta(x,y,t),
\label{36}
\\
\Psi^{(+)}\to -v_0z, \qquad z\to-\infty.
\label{37}
\end{gather}
The same equations can be immediately obtained from
Eqs.~(\ref{23})--(\ref{28}) in the limit $\rho\to 0$. They coincide with
the equations describing the so-called Laplacian growth, viz., the motion
of the phase boundary with a velocity directly proportional to the normal
derivative of a certain harmonic scalar field ($\Psi^{(+)}$ in our case).
Depending on the chosen frame of reference, this field may have the
meaning of temperature (Stefan's problem in the quasi-stationary limit),
electrostatic potential (electrolytic deposition), or pressure (flow
through a porous medium). It is important for us that there are many known
exact solutions to Eqs.~(\ref{34})--(\ref{37}). They describe the
evolution of the interface up to the formation of ``fingers'', cuspidal
dimples, and so on (see, for example, \cite{lap1,lap2,lap3,lap4}).

Let us prove that the class of solutions of the motion equations
(\ref{29})--(\ref{33}) corresponding to the reduced
Eqs.~(\ref{34})--(\ref{37}) is stable to small perturbations of potential
$\Psi^{(-)}$. It should be noted that the motion of the liquid-liquid
boundary described by Eqs.~(\ref{34})--(\ref{37}) is always directed
towards the lower liquid; this is associated with the extremum principle
for harmonic functions. Let function $\eta$ at the initial
instant $t_0$ be a single-valued function of variables $x$ and $y$. In
this case, for $t>t_0$, the inequality
\begin{equation}
\eta(x,y,t)\leq\eta(x,y,t_0)
\label{38}
\end{equation}
holds for any $x$ and $y$. This inequality remains valid for small
perturbations of $\Psi^{(-)}$ also, when the effect of potential
$\Psi^{(-)}$ in relation (\ref{31}) can be disregarded as compared to
the effect of potential $\Psi^{(+)}$, and the motion of the boundary
is described by the same Eqs.~(\ref{34})--(\ref{37}).

As regards the evolution of potential $\Psi^{(-)}$, it is sufficient, for
small $|\nabla\Psi^{(-)}|$, to consider the boundary condition
(\ref{30}) in the linear approximation. It takes the trivial form:
\begin{equation}
\Psi^{(-)}_t=0, \qquad z=\eta(x,y,t).
\label{39}
\end{equation}
This means that the potential does not change with time in the chosen
reference frame (the origin moves relative to the interface
with speed $v_0$). In the simplest case of a periodic perturbation, the
solution to Eqs.~(\ref{29}), (\ref{33}) and (\ref{39}) is given by
$$
\Psi^{(-)}=Ae^{\kappa z}\sin{(\kappa x)},
$$
where $\kappa$ is the perturbation wave number, and $A$ is a constant small
amplitude. Let us denote the potential at the boundary $z=\eta$ by $\psi$.
We have
$$
\psi(x,y,t)\equiv\Psi^{(-)}|_{z=\eta}=
Ae^{\kappa\eta(x,y,t)}\sin{(\kappa x)}.
$$
Taking into account the inequality (\ref{38}), we finally get
$$
\left|\psi(x,y,t)\right|\leq\left|\psi(x,y,t_0)\right|
$$
for any $x$ and $y$ at $t>t_0$, that is the value of the potential
$\Psi^{(-)}$ at the interface does not increase with time. Furthermore,
since the level of the interface (the value of function $\eta$
averaged over the spatial variables) moves downwards at a constant
velocity, it is evident that the potential $\Psi^{(-)}$ relaxes to zero at
the boundary. Thus, we have proved that Eqs.~(\ref{34})--(\ref{37})
describe the asymptotic behavior of the liquid-liquid interface in a
strong vertical electric field. 

It should be noted that the results of this work
can be used to describe the motion of the interface of two dielectric
liquids in an applied electric field for other geometries of the problem.
All one has to do is to modify the conditions (\ref{5}), (\ref{6}),
(\ref{11}), and (\ref{12}). This will allow us to consider the interface
dynamics in an oblique or tangential electric field, and also the
dynamics of closed interfaces in an external field.

In addition, the results of the above investigation can be extended to the
case of two magnetic fluids in a vertical magnetic field. For this
purpose one should replace the electric fields $E_{1,2}$ by the magnetic
fields $H_{1,2}$ and the permittivities $\varepsilon_{1,2}$ by the magnetic
permeabilities $\mu_{1,2}$. 

This study was supported by the ``Dynasty'' Foundation and the
International Center for Fundamental Physics in Moscow.

{}

\end{document}